\pdfoutput=1

\documentclass[10pt]{article} % For LaTeX2e
\usepackage[preprint]{tmlr}
\usepackage{natbib}

\usepackage{url}

\usepackage[breaklinks]{hyperref}

\hypersetup{breaklinks=true}

\usepackage{url}
\usepackage{CJKutf8}
\usepackage{graphicx}
\usepackage{longtable}
\usepackage{array}
\usepackage{booktabs}
\usepackage{multirow}
\usepackage[normalem]{ulem}
\usepackage{comment}
\usepackage[most]{tcolorbox}
\definecolor{whitesmoke}{HTML}{F5F5F5}
\colorlet{background}{whitesmoke}
\usepackage{enumitem}
\usepackage{float}
\usepackage{titlesec}
\titlespacing{\subsubsection}{4pt}{*0}{*0}

\usepackage{graphicx}
\usepackage{array}

\usepackage{array}
\usepackage{colortbl}

\makeatletter
\def\NAT@force@numbers{}
\makeatother

\makeatletter
\renewcommand\@biblabel[1]{}  % Removes the numbers
\makeatother

\newtcolorbox{researchbox}{%
breakable,boxrule=1pt,enhanced jigsaw, sharp corners,pad at break*=1mm,colbacktitle=background,colback=background,colframe=black,coltitle=black,toptitle=2mm,bottomtitle=1.5mm,width=\linewidth,fonttitle=\bfseries,parbox=false,title=Example Research
Questions,boxed title style={empty,boxrule=0pt, bottom=0pt}
}

\title{The Role of AI Safety Institutes in Contributing to International Standards for Frontier AI Safety}

\author{\name Kristina Fort \addr Pivotal Research \email kristina.s.fort@gmail.com \\ 
    }

\def\month{MM}  % Insert correct month for camera-ready version
\def\year{YYYY} % Insert correct year for camera-ready version
\def\openreview{\url{https://openreview.net/forum?id=XXXX}} % Insert correct link to OpenReview for camera-ready version

\begin{document}

\maketitle

\begin{abstract}
International standards are crucial for ensuring that frontier AI systems are developed and deployed safely around the world. Since the AI Safety Institutes (AISIs) possess in-house technical expertise, mandate for international engagement, and convening power in the national AI ecosystem while being a government institution, we argue that they are particularly well-positioned to contribute to the international standard-setting processes for AI safety. In this paper, we propose and evaluate three models for AISIs’ involvement: Model 1 – Seoul Declaration Signatories, Model 2 – US (+ other Seoul Declaration Signatories) and China, and Model 3 – Globally Inclusive. Leveraging their diverse strengths, these models are not mutually exclusive. Rather, they offer a multi-track system solution in which the central role of AISIs guarantees coherence among the different tracks and consistency in their AI safety focus.
\end{abstract}

% \makeatletter
% \let\origsection\section
% \renewcommand\section{\@ifstar{\starsection}{\nostarsection}}

% \newcommand\nostarsection[1]
% {\sectionprelude\origsection{#1}\sectionpostlude}

% \newcommand\starsection[1]
% {\sectionprelude\origsection*{#1}\sectionpostlude}

% \newcommand\sectionprelude{%
%   \vspace{1em}
% }

% \newcommand\sectionpostlude{%
%   \vspace{1em}
% }
% \makeatother

% \newpage
\section*{Executive Summary}
Standards, defined as “a variety of rules and norms in everyday use”\footnote{Christopher Thomas and Dr Florian Ostmann. “Enabling AI Governance and Innovation through Standards | UNESCO.” UNESCO, March 15, 2024. https://www.unesco.org/en/articles/enabling-ai-governance-and-innovation-through-standards.}, are widely seen as a necessary precondition for the safe development and deployment of frontier AI systems.\footnote{Following Anderljung et al. (2023), we define frontier AI systems as highly capable foundation models that could exhibit sufficiently dangerous capabilities.}\footnote{Markus Anderljung et al. “Frontier AI Regulation: Managing Emerging Risks to Public Safety.” arXiv, November 7, 2023. http://arxiv.org/abs/2307.03718.}

Because of their in-house technical expertise, focus on international engagement, and convening power while being a government institution, the AI Safety Institutes (AISIs) are well-placed to contribute to international standard-setting processes for AI safety.

We describe three potential “models” for cooperation on safety standards among AISIs and similar institutions. Each model brings together different sets of countries and will accordingly have different strengths and weaknesses. As a result, different models will be appropriate for different kinds of international standards. Our findings are summarized in the table below.

\textbf{Model 1: Seoul Declaration Signatories} consists of the original Seoul Declaration signatories – Australia, Canada, the European Union, France, Germany, Italy, Japan, the Republic of Korea, the Republic of Singapore, the United Kingdom, and the United States of America. 

\textbf{Model 2: The US (+ other Seoul Declaration Signatories) and China} is a model targeted at reaching a consensus between the US and China and might include the original Seoul Declaration signatories as well.

\textbf{Model 3: Globally Inclusive} reflects the current membership of the ISO/IEC sub-committee for Artificial Intelligence (JTC 1/SC 42), which assembles a broad range of countries, such as Germany, Egypt, Brazil, Canada, and India.

The various models are not mutually exclusive. Each of them might have an important role to play in the international standard-setting ecosystem, with the involvement of the AISIs across the various tracks adding consistency and coherence to the ecosystem.

\begin{table}[htbp]
\caption{Summary of the three proposed models for the involvement of AISIs.}
\centering
\begin{tabular}{|>{\columncolor[HTML]{DAE8FC}}p{4cm}|c|c|c|p{4cm}|}
\hline
\cellcolor[HTML]{DAE8FC}\textbf{Model}                              & \cellcolor[HTML]{D5E8D4}\textbf{Responsiveness} & \cellcolor[HTML]{D5E8D4}\textbf{Legitimacy} & \cellcolor[HTML]{D5E8D4}\textbf{Expertise} & \cellcolor[HTML]{D5E8D4}\textbf{Examples of Suitable Standards} \\ \hline
1. Seoul Declaration Signatories    & High    & Low   & High  & Best suited for standards that reflect the risk appetite and values of like-minded actors, such as process and management standards and product testing, performance and measurement standards.
\underline{Example:} safety thresholds for frontier AI modelss \\ \hline
2. US (+ Seoul Declaration Signatories) and China & Medium  & Medium & High  & Best suited for standards that foster mutual trust and establish a safety baseline, such as process and management standards. \underline{Example:} incident response and recovery plans \\ \hline
3. Globally Inclusive & Low  & High & Low  & Best suited for developing standards that benefit from as broad a participation as possible, such as interface and networking standards.
\underline{Example:} shared methods for interpretability \\ \hline
\end{tabular}
\end{table}

\subsection*{Acknowledgements}
We are particularly grateful to Oliver Guest for his mentorship, guidance, and valuable feedback on this research project. We are grateful to Belinda Cleeland, Hadrien Pouget and Peter Cihon for sharing their insights and observations on this topic. All mistakes remain our own.

\newpage
\tableofcontents

\section{Introduction}
International standard-setting is crucial for developing usable and harmonized international AI standards that contribute to the safe development and deployment of frontier AI systems.\footnote{Following Anderljung et al. (2023), we define frontier AI systems as highly capable foundation models that could exhibit sufficiently dangerous capabilities.} The International Organization for Standardization (ISO) defines standards as “a formula that describes the best way of doing something.”\footnote{ ISO. “ISO - Standards.” ISO. Accessed August 14, 2024. https://www.iso.org/standards.html.} Other authors describe them as “a variety of rules and norms in everyday use.”\footnote{Christopher Thomas and Dr Florian Ostmann. “Enabling AI Governance and Innovation through Standards | UNESCO.” UNESCO, March 15, 2024. https://www.unesco.org/en/articles/enabling-ai-governance-and-innovation-through-standards.} Standards are seen as one of the key means to “enable good governance practices by establishing consistent measurement and norms for interoperability”.\footnote{Mina Narayanan et al. “Repurposing the Wheel: Lessons for AI Standards.” Center for Security and Emerging Technology. Accessed August 2, 2024. https://cset.georgetown.edu/publication/repurposing-the-wheel/.
} They have the potential to contribute to frontier AI safety by operationalizing AI regulation, providing clear and coherent guidelines to the industry, and setting a comparable baseline for evaluations.\footnote{Renan Araujo, Kristina Fort, and Oliver Guest. “Understanding AI Safety Institutes: Core Characteristics, Functions, and Challenges.” Institute for AI Policy and Strategy, 2024 [forthcoming].
}

While there seems to be a consensus on the general importance of standards for ensuring frontier AI safety, it is often unclear what is meant by the term “standards” and what type of standards is needed to achieve the aforementioned goals.\footnote{Tim McGarr. “Standards as a Foundation for Success with AI.” AI Standards Hub, December 6, 2023. https://aistandardshub.org/standards-as-a-foundation-for-success-with-ai/.
} International standardization efforts around AI have been concentrating mainly on “standards to improve market efficiency and address ethical concerns,”\footnote{Peter Cihon. “Standards for AI Governance: International Standards to Enable Global Coordination in AI Research \& Development.” GovAI, April 17, 2019. https://www.governance.ai/research-paper/standards-for-ai-governance-international-standards-to-enable-global-coordination-in-ai-research-development. } but these may prove insufficient “to address further policy objectives, such as a culture of responsible deployment and use of safety specifications in fundamental research.”\footnote{ Ibid.} Therefore, it is crucial to define what type of AI standards are currently lacking and align them closely with AI policy objectives.

AI Safety Institutes (AISIs) might have an important role to play in this. AISIs are “safety-focused, state-backed, technocratic, domestic institutions that perform several different functions”, including standard-setting.\footnote{Renan Araujo, Kristina Fort, and Oliver Guest. “Understanding AI Safety Institutes: Core Characteristics, Functions, and Challenges.” Institute for AI Policy and Strategy, 2024 [forthcoming].
} As of August 2024, AISIs have been established in the UK, the USA, Japan, Singapore, and Canada, and the EU AI Office is often considered an “AISI-like” institution. While their engagement in standard-setting has been fairly limited so far, the US AISI is officially tasked with facilitating “the development of trusted standards” and the EU AI Office is overseeing the development of standards led by CEN-CENELEC (European Committee for Electrotechnical Standardization) that should lay a foundation for the EU AI Act.\footnote{NIST. “NIST Risk Management Framework Aims to Improve Trustworthiness of Artificial Intelligence.” NIST, January 26, 2023. https://www.nist.gov/news-events/news/2023/01/nist-risk-management-framework-aims-improve-trustworthiness-artificial.
} 

The AI Safety Institutes have three features that provide them with a great foundation for participation in international standard-setting: 

Firstly, they have in-house technical expertise while being a government institution, which allows them to bring expert opinions to the debate and to conduct foundational scientific research supporting standard-setting. At the same time, their government association gives them credibility to contribute to standard-setting. 

Secondly, they are already active internationally, establishing different forms of cooperation with each other, and it is reasonable to expect that this cooperation will deepen through an international network of AISIs proposed at the Seoul AI Summit.\footnote{NIST. “U.S. Secretary of Commerce Gina Raimondo Releases Strategic Vision on AI Safety, Announces Plan for Global Cooperation Among AI Safety Institutes.” NIST, May 21, 2024. https://www.nist.gov/news-events/news/2024/05/us-secretary-commerce-gina-raimondo-releases-strategic-vision-ai-safety.
}

Finally, their convening power enables the AISIs to contribute to international standard-setting, reflecting the opinions of the domestic AI ecosystem and representing the public interest. Therefore, the AISIs have a role to play in contributing to frontier AI safety-relevant standards.

In this policy brief, we aim to:
\begin{enumerate}
\item Introduce the international standard-setting landscape,
\item Outline the types of standards relevant to AI standardization, and
\item Present three potential models of cooperation among the AI Safety Institutes (AISIs) and other standards development organizations (SDOs) to advance AI safety standard-setting efforts internationally.
\end{enumerate}

\section{International Standardization for AI}

\subsection{International Standards and Their Taxonomy}

Standards generally represent a package of useful harmonization guidelines but opinions on what specific standards could contribute to the safety of frontier AI systems differ. This section introduces standard taxonomy and clarifies which types of standards are important for frontier AI safety. 

When discussing standards in the context of AI, people often mean different things. Therefore, it is important to define and classify standards and we will try to do this through the introduction of the following taxonomy, described by Ostmann and Thomas. This taxonomy revolves around the purpose of standards,\footnote{Christopher Thomas and Dr Florian Ostmann. “Enabling AI Governance and Innovation through Standards | UNESCO.” UNESCO, March 15, 2024. https://www.unesco.org/en/articles/enabling-ai-governance-and-innovation-through-standards.
} dividing them into five categories:
\begin{enumerate}
    \item foundational and terminology standards
\item process and management standards
\item measurement standards
\item product testing and performance standards, and 
\item interface and networking standards.\footnote{ Christopher Thomas and Dr Florian Ostmann. “Enabling AI Governance and Innovation through Standards | UNESCO.” UNESCO, March 15, 2024. https://www.unesco.org/en/articles/enabling-ai-governance-and-innovation-through-standards.
} 
\end{enumerate}

For our purposes, we will group together measurement standards and product testing and performance standards as they overlap and complement each other to a large extent when it comes to frontier AI safety.\footnote{Rosamund Powell et al. “Towards Secure AI.” Centre for Emerging Technology and Security, March 2024. https://cetas.turing.ac.uk/publications/towards-secure-ai.
} 

In the sections dedicated to the individual models, we consider which models of international standards collaboration would be the best fit for each type of the aforementioned standards. We also aim to take their importance for frontier AI safety and maturity into account.\footnote{Adapted from Rosamund Powell et al. “Towards Secure AI.” Centre for Emerging Technology and Security, March 2024. https://cetas.turing.ac.uk/publications/towards-secure-ai.} The importance for frontier AI safety is measured in terms of the expected direct positive impact a given standard could have on frontier AI safety. In terms of maturity, we consider if the state-of-the-art scientific research provides enough evidence to standardize and potentially enforce a given standard.

\subsection{An Overview of International Standard-Setting Organizations for AI}

International standard-setting processes are complex and murky. The standardization landscape is highly complicated and decentralized with many different organizations engaged in the standardization process. The main standard-setting players are usually industry consortia and national, regional, and international standards bodies, which mostly share two main features: their decisions are “industry-led and consensus-driven” and their standards are “voluntary or non-binding.”\footnote{Matt Sheehan and Jacob Feldgoise. “What Washington Gets Wrong About China and Technical Standards,” February 27, 2023. https://carnegieendowment.orgundefined?lang=en.
} While standards are mostly voluntary or non-binding, they establish criteria or describe processes that can become binding, if policymakers reference them in regulation. 

The largest and most well-established technical standards development organizations (SDOs) in the field of AI are the International Organization for Standardization (ISO) and the International Electrotechnical Commission (IEC), which created a joint standards sub-committee for Artificial Intelligence (JTC 1/SC 42) that has been actively working on developing various technical AI standards. These organizations are private institutions which assemble national delegations from many different countries with equal vote. They cooperate closely with other standard-setting bodies, such as the EU's CEN-CENELEC, through various agreements.

There are also other intergovernmental or independent SDOs engaged in AI standardization. The Institute of Electrical and Electronics Engineers (IEEE) has a Standards Association which has worked on developing AI standards as well and, along with the ISO and the IEC, is deemed to occupy a firm position in this field.\footnote{Hans W. Micklitz. “The Role of Standards in Future EU Digital Policy Legislation.” ANEC, July 2023. Accessed August 9, 2024. https://www.anec.eu/images/Publications/other-publications/2023/ANEC-DIGITAL-2023-G-138.pdf.} However, the IEEE is not truly international since the participating members are disproportionately US-based stakeholders. The International Telecommunication Union (ITU) has also been aiming to gain a stronger footing in AI standard-setting through pre-standardization focus groups and an annual AI summit but its position within the United Nations ecosystem and its more politicized context complicates its engagement in this domain. 

The ITU has been criticised for “having insufficiently rigorous procedures and protections, leading to low quality or politically-motivated proposals” which has led to the drop out of several important actors, such as some major US companies.\footnote{Matt Sheehan and Jacob Feldgoise. “What Washington Gets Wrong About China and Technical Standards,” February 27, 2023. https://carnegieendowment.org/research/2023/02/what-washington-gets-wrong-about-china-and-technical-standards?lang=en.
} Therefore, among the SDOs, the ISO/IEC tandem may be more promising than the ITU for creating AI standards. As independent, non-governmental organizations that unite national standards bodies (ISO) and national committees (IEC), the ISO and the IEC are in a unique position. Moreover, they appear to be successful in leveraging this position to facilitate meaningful collaboration on technical AI standards.\footnote{Nora von Ingersleben-Seip. “Competition and Cooperation in Artificial Intelligence Standard Setting: Explaining Emergent Patterns.” Review of Policy Research 40, no. 5 (January 25, 2023): 781–810. https://doi.org/10.1111/ropr.12538.
} 

The ISO/IEC JTC 1/SC 42 has managed to publish several documents relevant to AI standardization, defining key concepts and terminology, assessing the robustness of neural networks, and offering guidance on risk management.\footnote{International Standards Organization. “ISO/IEC JTC 1/SC 42 - Artificial Intelligence.” Accessed August 5, 2024. https://www.iso.org/committee/6794475/x/catalogue/p/1/u/0/w/0/d/0.} Therefore, when considering existing international AI standardization bodies, this policy brief will primarily focus on the ISO/IEC SC 42, as it is the most active and well-established actor in international standard-setting.

At the same time, the standardization process led by the ISO/IEC SC 42 faces some key limitations. The ISO and IEC have a constrained playing field as they do not set their own agenda. Instead, they react to the market need, which is demonstrated through the approved proposals from their members. Even more importantly, the standardization process is quite slow which is particularly problematic for AI standard-setting as AI development progresses rapidly. Furthermore, the revision of standards might cause delays as, typically, standards are revised every five years. However, this process depends directly on the decision of the committee which can theoretically launch the revision procedure at any time it chooses. 

One final consideration for the ISO/IEC process is the availability of standards. Since the ISO and the IEC are independent organizations, they gain funding (royalties) through publishing and providing access to their standards which are mostly behind a paywall. Therefore, if standards are viewed as a public good that should be accessible to anyone, the ISO/IEC model fails to satisfy this need, restricting its benefit-sharing potential. Greater involvement of AISIs, actors external to the ISO/IEC process, could therefore expand democratisation and benefit-sharing in international standard-setting, since the financial existence of AISIs is independent of such royalties.

\subsection{Current Standardization Efforts}

So far, Standards Development Organizations (SDOs) and other entities have primarily concentrated on foundational and terminology standards to build a common language around AI, dedicating their attention to general AI standards rather than those specifically tailored to frontier AI safety. Recently, the focus has begun to shift towards process and management standards as the ISO/IEC issued their guidance on risk management and the National Institute of Standards and Technology (NIST) published their AI Risk Management Framework (AI RMF). While there are some standards for the remaining two categories, more work is necessary in this domain.  

NIST prepared “A Plan for Engagement on AI Standards”, which introduces several topics as ready for standardization and urgent.\footnote{NIST. “A Plan for Global Engagement on AI Standards.” Gaithersburg, MD: National Institute of Standards and Technology, 2024. https://doi.org/10.6028/NIST.AI.100-5.} Apart from terminology and taxonomy, these include “shared testing, evaluation, verification, and validation (TEVV) practices for AI models and systems”, “mechanisms for enhancing awareness and transparency about the origins of digital content”, transparency around data characteristics and practices, and “incident response and recovery plans.”\footnote{Ibid.} NIST’s Plan also lists topics that might be important but require further scientific foundational work, such as interpretability and explainability techniques, human-AI configuration, and AI resource consumption.\footnote{Ibid.}

This list of standardization topics compiled by NIST serves as a useful illustration of what standards are necessary for frontier AI safety. Additional topics for frontier AI standardization might include “engaging external experts for independent scrutiny,” “standardized protocols for frontier AI model deployment,” and “monitoring of new information on model capabilities.”\footnote{Markus Anderljung et al. “Frontier AI Regulation: Managing Emerging Risks to Public Safety.” arXiv, November 7, 2023. http://arxiv.org/abs/2307.03718.} These standards, requiring technical as well as non-technical considerations for decision-making, seem to be particularly relevant for the AISIs. Especially since some AISIs, such as the UK AISI, currently perform evaluations of AI models which provide them with important scientific insights useful for standardization.\footnote{Renan Araujo, Kristina Fort, and Oliver Guest. “Understanding AI Safety Institutes: Core Characteristics, Functions, and Challenges.” Institute for AI Policy and Strategy, 2024 [forthcoming].}  

\section{Criteria for International Standard-Setting Models}

To judge the different models of AISI involvement in international standardization, we need to identify the characteristics that international standard-setting processes would ideally have. This section describes the three main characteristics against which we evaluate the different models: responsiveness, legitimacy, and expertise.

\textbf{Responsiveness:} AI progress is very fast and fairly unpredictable which requires agility and adaptability from the SDOs. This translates to pacing the standardization process so it can keep up with the rapid progress in AI development. The SDOs need to remain responsive to changes in the field and revise standards to keep them relevant even if AI paradigms shift.

\textbf{Legitimacy:} Standards must be trusted, meaning that a broad range of stakeholders must be involved in the standardization process. SDOs need to balance broad stakeholder engagement with risks from industry capture to give the process legitimacy. In addition, all the relevant countries need to endorse the standards to make them legitimate, which is connected with them being involved or represented in the SDO.

\textbf{Expertise:} Finally, legitimacy and responsiveness are closely connected to technical expertise. Technically feasible and specific standards are crucial for ensuring objective assessment of compliance and enforcement. Therefore, technical expertise must be present in the SDOs to make the standards usable and operational.

SDOs need to find a balance between these criteria and navigate potential tradeoffs. For example, they must ensure a speedy process while involving a large number of stakeholders, which can sound quite paradoxical. Fully acknowledging this fuzziness, they still offer a valuable benchmark for evaluating the potential of our three models for AISI involvement in the AI standardization efforts.

\section{Models}

\subsection{Model 1: Seoul Declaration Signatories}

The first model gives the AI Safety Institutes the greatest agency in international standard-setting. This setup builds on the Seoul Declaration, which includes a commitment to “promote the development and adoption of international standards”\footnote{UK Government. “Seoul Statement of Intent toward International Cooperation on AI Safety Science, AI Seoul Summit 2024 (Annex).” GOV.UK, May 21, 2024. https://www.gov.uk/government/publications/seoul-declaration-for-safe-innovative-and-inclusive-ai-ai-seoul-summit-2024/seoul-statement-of-intent-toward-international-cooperation-on-ai-safety-science-ai-seoul-summit-2024-annex.} and unites countries that are geopolitically reasonably aligned with each other. In this model, the AI Safety Institutes and “AISI-like institutions” in the original Seoul Declaration signatory countries should join forces through the International Network of AISIs to set harmonized standards for themselves.

Members in this standardization model would be the currently existing AISIs and AISI-like institutions from Australia, Canada, the European Union, France, Germany, Italy, Japan, the Republic of Korea, the Republic of Singapore, the United Kingdom, and the United States of America.

\textbf{Responsiveness:} Uniting the AISIs and AISI-like institutions in the US-allied democracies, mostly G7 states, to work on AI standardization would likely make the standardization process quicker and more responsive. As a small group of fairly like-minded countries, they could accelerate the process and also revise their outputs regularly. 

\textbf{Legitimacy:} On the one hand, the exclusivity of this model might slightly decrease its legitimacy since it would be a mainly rich developed nations’ initiative. Therefore, the standards developed in this model might face resistance from other countries, especially China, viewing the process as illegitimate, as well as pushback from the traditional SDOs, losing their prime position in international standard-setting. 

On the other hand, the convening power of AISIs and their important role in the national and international AI ecosystems might increase the legitimacy of this process, at least in the allied countries. While the global reach of this initiative might appear limited, the AISI standards, leveraging the AISI reputation, could definitely serve as an inspiration for other countries, leading to a possible version of the Brussels effect that we call “the AISI effect,” especially if the AISI-developed AI standards become binding in the involved countries.

\textbf{Expertise:} The AISIs would bring in their high technical expertise, which could solidify trust in this standardization process. Furthermore, their foundational scientific research could directly feed into the international standard-setting and a more active coordination among the AISIs could allow them to cover more complex scientific topics relevant to standardization.

\textbf{Feasibility:} To make this model possible, the mandate of most AISIs as well as their resourcing would need to expand. While the US AISI has a standard-setting angle, most AISIs and AISI-like institutions, most importantly the UK AISI, would need to have their mandate re-defined or expanded to assume an AI standardization role. The UK AISI is currently an international leader in AI safety and its greater engagement in international standard-setting could meaningfully steer this model. The expansion of the mandate would need to go hand in hand with boosting their staff capacities to keep up with this new stream of responsibilities and increasing their expertise in standardization. Since AISIs are state-backed institutions, the potential development of their responsibilities would depend mainly on political vision and will. Especially as the ambitions set for the AISIs vary a lot between countries and some of the Seoul Declaration signatories have not established their AISIs yet.

\textbf{Fit:} This model might be particularly good for developing standards for frontier AI safety in two domains: process and management standards, and product testing, performance, and measurement standards. Since all the engaged countries highly value the safety and protection of human rights and civil liberties, establishing processes for responsible development and deployment of advanced AI systems, agreeing on common safety thresholds and metrics, and creating shared TEVV practices would be particularly valuable. Moreover, this seems to be more achievable in a group of like-minded partners. Standards developed in this model could be later submitted to the ISO/IEC JTC 1/SC 42 to transform them into international standards.

\subsection{Model 2: US (+ Seoul Declaration Signatories) and China}

The second model builds on the AI Safety Institutes’ engagement in dissemination and diplomacy efforts, expanding the first standardization model by adding China to the mix of involved actors.\footnote{Renan Araujo, Kristina Fort, and Oliver Guest. “Understanding AI Safety Institutes: Core Characteristics, Functions, and Challenges.” Institute for AI Policy and Strategy, 2024 [forthcoming].} Therefore, the process would be focused mainly on standards contributing to mutual trust for the advancement of AI safety. In this model, the AISIs would assume stronger standard-setting and diplomatic responsibilities, negotiating AI standards between themselves and China. China does not formally have an AISI, though it does have institutions that are in some ways analogous, such as the China Academy for Information and Communications Technology (CAICT) and the Beijing Academy of Artificial Intelligence (BAAI).\footnote{Oliver Guest \& Karson Elmgren. “Chinese AISI Counterparts.” 2024 [forthcoming].} China could also set up a body that more closely resembles an AISI; rumors have been circulating that this might happen.\footnote{Matt Sheehan. “China’s Views on AI Safety Are Changing—Quickly.” Carnegie Endowment for International Peace, August 27, 2024. https://carnegieendowment.org/research/2024/08/china-artificial-intelligence-ai-safety-regulation?lang=en.
}

China has been active in international standardization efforts, participating in many of the outlined SDOs. Moreover, it has started its own domestic AI standardization process through the CAICT, focusing on the standardization of “recommendation algorithms, deepfakes, and content security,” but also providing a draft standard on “specific tests a generative AI model must pass before registering”.\footnote{Matt Sheehan. “Tracing the Roots of China’s AI Regulations.” Carnegie Endowment for International Peace, February 27, 2024. https://carnegieendowment.orgundefined?lang=en.} China has also shown interest in AI safety, attending the AI Safety Summit in the United Kingdom and the AI Summit in South Korea and participating in the Geneva talks as well as Track 2 dialogues with the United States. Therefore, standard-setting negotiations could happen between the AISIs and their closest counterpart in China to lay down common foundations for regulation in “the West” and in China.

\textbf{Responsiveness:} This model holds the promise of being quicker and more responsive than the traditional standardization process since both sides are very active in AI development and deployment and common standards could contribute to mitigating a dangerous competitive dynamic between them. Furthermore, this process could be more responsive to alleviating grievances particular to the United States and China with other involved actors serving as mediators. On the other hand, China might understandably be reluctant to take part in a forum that is primarily composed of a group of countries that are more aligned with the United States. Therefore, this model could also take shape through cooperation between the US AISI and a Chinese AISI-like institution only to preserve its responsiveness.

\textbf{Legitimacy:} While the process would likely take place behind closed doors, it is reasonable to expect that both the US AISI and the Chinese representative would be familiar with the interests of their domestic ecosystem thanks to their convening power. This process could appear less legitimate as it mostly excludes third actors, such as industry and civil society, as well as the Global Majority. Still, some level of “the AISI effect” might occur since standards supported by the largest AI actors would likely diffuse internationally, especially when they are developed by top technical experts from the involved countries.

\textbf{Expertise:} As both the US and China are largely viewed as AI powerhouses, they could build on their high technical expertise. Moreover, since both countries are highly engaged in international standardization procedures, such as in the ISO/IEC and the ITU, they have countless experts with a solid understanding of international standard-setting who could join the AISIs to steer the process forward. Therefore, they are well-prepared to establish bilateral standards.

\textbf{Feasibility:} Bringing this model to life would require strong political backing and diplomatic engagement that could materialize in preparation for the AI Action Summit in France. The main barrier to the success of this model is likely the US suspicion of Chinese involvement in international standardization since “the Chinese government has a track record of trying to manipulate international organizations” and it has used its Belt and Road Initiative to push for more favorable standards for China.\footnote{Matt Sheehan and Jacob Feldgoise. “What Washington Gets Wrong About China and Technical Standards,” February 27, 2023. https://carnegieendowment.orgundefined?lang=en.
} However, as the model is trying to avoid this suspicion by engaging with China in terms of international standardization in a smaller and less diffuse forum, this might be the right strategy to create a level playing field and negotiate relevant AI safety standards.

\textbf{Fit:} This model might be useful especially for developing process and management standards since it could contribute to greater frontier AI safety in the leading companies in the US and China. Policy standards, such as incident response and recovery plans, seem to be particularly valuable as these might also facilitate greater mutual trust between the two countries and establish an AI safety baseline for their AI development and deployment.

\subsection{Model 3: Globally Inclusive}

The third and final model considers AISIs as potential contributors to the AI standardization process in the traditional SDOs, specifically the ISO/IEC joint standards sub-committee 42 (JTC 1/SC 42). In this model, AISIs would have the least active role as the standardization process in ISO/IEC is already well-established and codified. 

Each country can only have one representing institution, usually one national standardization body, in the ISO, which means that the AISIs would have to adopt the role of advisors to their official representative on AI-related standardization matters and potentially get a nomination for one of their experts to represent their country in the JTC 1/SC 42’s technical working group. Engagement in the working group should be purely technical, building scientific consensus and disregarding national interests. 

Currently, the JTC 1/SC 42 has 39 participating members and 26 observing members, with all jurisdictions that have an AISI being participating members already, except for the EU.\footnote{International Standards Organization. “ISO/IEC JTC 1/SC 42 - Artificial Intelligence.” Accessed August 9, 2024. https://www.iso.org/committee/6794475/x/catalogue/p/1/u/0/w/0/d/0.} Therefore, AISIs would cooperate with their national standardization organizations and with each other to boost international efforts in standard-setting. This would also involve working with various countries that do not have an AI Safety Institute (yet), such as Rwanda, Saudi Arabia or the Philippines.

\textbf{Responsiveness:} As outlined in the introduction, the standardization process in ISO/IEC is relatively slow, which means that this model is likely to fail to catch up with the rapid progress in AI development. This might be exacerbated by the revision process which takes place only every five years – this seems to be a very long time period when compared to current AI timelines.

\textbf{Legitimacy:} The ISO is a trusted organization that has created more than 24,500 standards across many industries and, therefore, is largely viewed as the main standardization body building global consensus.\footnote{Trager, Robert et al. “International Governance of Civilian AI: A Jurisdictional Certification Approach.” SSRN Scholarly Paper. Rochester, NY, September 23, 2023. https://doi.org/10.2139/ssrn.4579899.
} This gives the ISO/IEC JTC 1/SC 42 great legitimacy and expertise in how a regular standardization process works. In addition, the active involvement of Global Majority countries, such as Brazil, India, and Saudi Arabia, makes this standard process more legitimate in the sense of broad participation.

\textbf{Expertise:} The members of the ISO/IEC JTC 1/SC 42 mostly lack deep technical expertise in AI (compared to AISIs) as they are general standardization organizations. Instead, they are tasked with engaging with domestic stakeholders to represent their positions. This could lead to industry capture which could be prevented if AISIs become their main advisors on AI-related questions, representing the public interest while providing relevant technical insights.

\textbf{Feasibility:} This model seems to be the easiest to operationalize as the ISO/IEC JTC 1/SC 42 currently has 31 standards on AI already published and 35 under development.\footnote{International Standards Organization. “ISO/IEC JTC 1/SC 42 - Artificial Intelligence.” Accessed August 9, 2024. https://www.iso.org/committee/6794475/x/catalogue/p/1/u/0/w/0/d/0.
} However, the engagement of AISIs in this process has largely been omitted so far. Therefore, their mandate might need to expand to cover involvement in the standardization discussions through interactions with the domestic standardization bodies.

\textbf{Fit:} This model might be particularly useful for developing two types of standards: foundational and terminology standards and interface and networking standards. Foundational and terminology standards are crucial to establishing a common vocabulary, which can serve as a foundation for the development of other types of standards. Therefore, SDOs and other bodies can build on these in their standardization work. The development of interface and networking standards in the ISO/IEC context seems to be particularly important as these standards guarantee that frontier AI systems will be interoperable internationally, which requires broad participation and adoption.

\section{Conclusion}

This policy brief aimed to outline the international standard-setting landscape and propose three different models for expanding and accelerating the international frontier AI standardization process. We believe that the AISIs have a role to play in AI standardization. At the same time, we acknowledge that this role is placed on a spectrum – AISIs can conduct foundational scientific research supporting standardization efforts and provide advice to the SDOs, but they also have very good foundations to act as SDOs themselves under the right circumstances. 

Our three models are not mutually exclusive. Rather, they come with different advantages and disadvantages and, therefore, their unique position can be leveraged to contribute to various standard categories. We therefore believe that international AI standardization could take place in the form of a multi-track system. The central role of AISIs in this system could ensure coherence among the different tracks and consistency in the AI safety focus, preventing further fragmentation.

% \section*{Bibliography}  % Use \section* to avoid numbering the section

\end{document}